\documentclass{edp-conf}
\usepackage{graphicx}

\usepackage{amsmath}
\renewcommand{\d}{\ensuremath{\,\mathrm{d}}} 
\newcommand{\vecr}{\ensuremath{\mathbf{r}}}
\newcommand{\vecv}{\ensuremath{\mathbf{v}}}
\newcommand{\calL}{\ensuremath{\mathcal L}}
\begin{document}

\TitreGlobal{SF2A 2001}

\title{Kinemetry: quantifying kinematic maps}
\runningtitle{Kinemetry}
\author{Y.\ Copin}
\address{Sterrewacht Leiden, The Netherlands, \email{ycopin@strw.leidenuniv.nl}}
\author{R.\ Bacon}
\address{CRAL-Observatoire de Lyon, France} 
\author{M.\ Bureau$^{1}$}
\author{R.L.\ Davies}
\address{University of Durham, UK}
\author{E.\ Emsellem$^{2}$}
\author{H.\ Kuntschner$^{3}$}
\author{B.\ Miller}
\address{International Gemini Observatory, Chile}
\author{R.\ Peletier}
\address{University of Nottingham, UK}
\author{E.K.\ Verolme$^{1}$}
\author{P.T.\ de Zeeuw$^{1}$}
\maketitle
\begin{abstract} 
We describe a new technique, \emph{kinemetry}, to quantify kinematic
maps of early-type galaxies in an efficient way.  We present the first
applications to velocity fields obtained with the integral-field
spectrograph \textsc{sauron}.
\end{abstract}
%

Integral-field spectrographs (IFS) were designed to provide spectra over the
full spatial extent of the observed objects. Applied to early-type galaxies,
they can produce complete maps of the stellar kinematics. To date, these maps
have been used `as is', without fully exploiting all the information they
contain (e.g., Davies \etal\ 2001). Still, it is known from the study of the
surface brightness distribution of early-type galaxies that proper
quantification, e.g., via the $a_{4}$ distortion parameter, can shed new light
on these objects (Kormendy \& Djorgovski 1989). For this reason we have
developed \emph{kinemetry}, a new formalism which extends the photometric
approach to kinematic maps obtained by IFS.\looseness=-2

\section{Photometry \& kinemetry}
\label{sec:phot-kin}

The core of the dynamics of an early-type galaxy is embodied by its
stellar distribution function (DF) $f(\vecr \equiv (x,y,z),\vecv
\equiv (v_{x},v_{y},v_{[z]}))$. This fundamental quantity cannot be
measured directly, but different observations constrain it.

The surface brightness at any point in a galaxy image equals the DF
integrated along the line-of-sight (LOS): $\mu(x,y) =
\int_{\mathrm{LOS}}\!\d z \iiint \!\d\vecv f(\vecr,\vecv)$. The study
of this projected light distribution is often referred to as the
\emph{surface photometry}.

Spectroscopic observations of a galaxy provide the line-of-sight velocity
distribution (LOSVD): $\calL(v;x,y) = \int_{\mathrm{LOS}}\!\d z\iint\!\d
v_{x}\d v_{y}\,f(\vecr,\vecv)$. This is the ultimate kinematic observation for
early-type galaxies. The surface brightness is just the zeroth-order moment of
the LOSVD. Its low-order moments $V$, $\sigma$, and related quantities $h_{3}$
and $h_{4}$, are in regular use as probes of the galactic DF (e.g., van der
Marel \& Franx 1993).

In what follows, we extend the techniques of surface photometry to the
stellar kinematic maps, i.e., to the higher projected moments of the
DF. This approach, which we christen \emph{kinemetry}, has never been
developed fully, mostly because the required maps were difficult to
obtain until the recent advent of panoramic IFS.

\section{Kinemetric expansion}
\label{sec:expansion}

Few papers have addressed the problem of parametrization of the
stellar kinematics of early-type galaxies (e.g., Franx \etal\ 1991,
Statler 1994), and none have considered the case of IFS
data. Following techniques developed for H\textsc{i} maps (e.g.,
Schoenmakers \etal\ 1997), we use a simple Fourier expansion of the
projected kinematic quantity $K(x,y)$ expressed in polar coordinates
on the plane of the sky:
$$
K(r,\theta) 
  = a_{0}(r) + \sum_{i=1}^{N} c_{i}(r)\,\cos[i(\theta-\phi_{i}(r))]. 
$$
This harmonic expansion is the most straightforward and powerful
choice, and nicely follows the usual surface photometry approach
(e.g., Franx \etal\ 1989). 

Here we consider only the mean velocity fields. Similar expansions can
be applied to higher moments, once the moment parity is properly taken
into account. A full description of the method will be published
elsewhere.

\section{Applications}
\label{sec:apps}

We now present some preliminary results of the kinemetry technique, as applied
to the \textsc{sauron} observations of elliptical galaxies (Bacon et al.\ 
2001, de Zeeuw \etal\ 2001, Copin \etal\ this volume).

\begin{figure}
  \begin{center}
    \includegraphics[width=0.48\textwidth]{copin2_fig1a.eps}
    \includegraphics[width=0.51\textwidth]{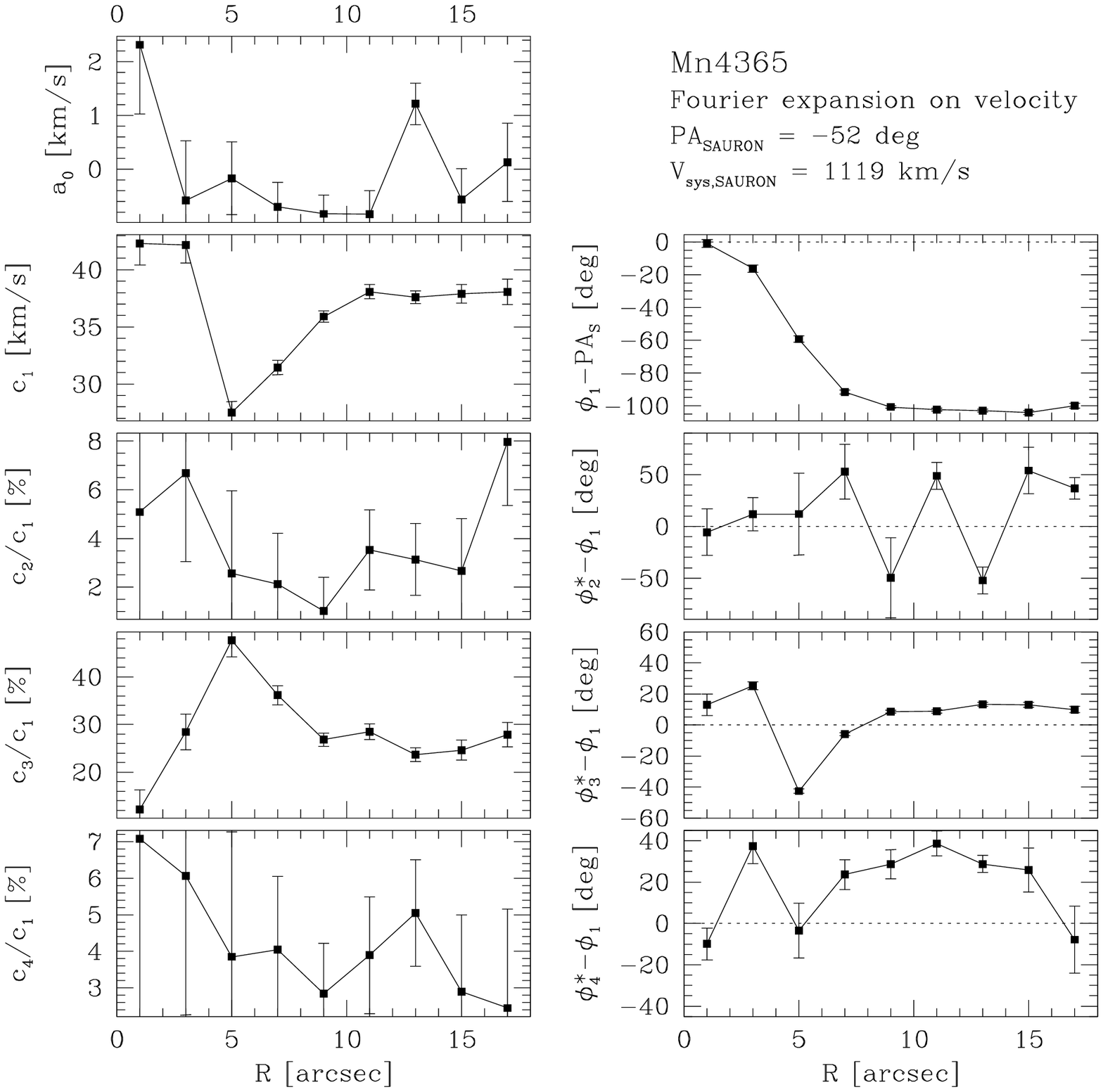}
    \includegraphics[width=0.48\textwidth]{copin2_fig1c.eps}
    \includegraphics[width=0.51\textwidth]{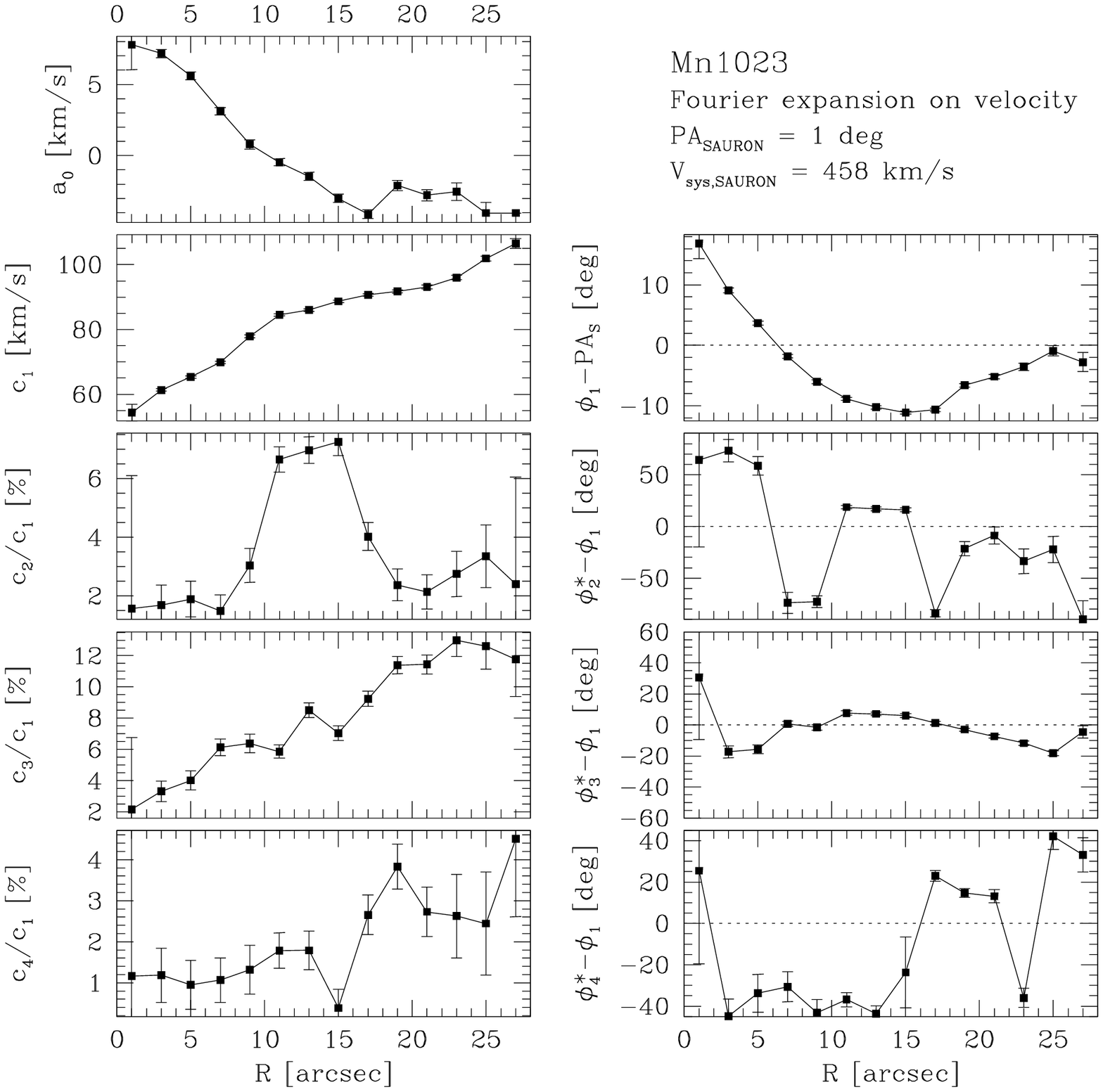}
    \end{center} 
\caption{Kinemetric analysis of NGC~4365 (\emph{upper panel}) and
  NGC~1023 (\emph{lower panel}). \emph{On the left:} observed velocity field
  (\emph{grayscale}), fourth-order reconstruction (\emph{white contours}) and
  `triaxial' (for NGC~4365) or `axisymmetric' (for NGC~1023) approximations
  (\emph{black contours}). \emph{On the right:} kinemetric parameters --
  amplitudes (\emph{left panel}) and phases (\emph{right panels}) up to fourth
  order -- as function of radius.}
\label{fig:kinemetry}
\end{figure}

\paragraph{Kinemetric parameters.}
\label{sec:param}
Fig.~\ref{fig:kinemetry} shows the dependence of the amplitudes
$c_{i}$ and phases $\phi_{i}$ on radius in the galaxies NGC~4365 (E3)
and NGC~1023 (SB0). The coefficient $c_{1}$ is always \emph{dominant},
and is related to the amplitude of the velocity curve. We therefore
express the $c_{i>1}$ as fractions of $c_{1}$. The phase $\phi_{1}$ is
the \emph{kinematic angle}, and gives the main orientation of the
velocity field at a given radius. Its radial dependence clearly
displays the extent and orientation of kinematically decoupled cores
(e.g., NGC~4365) and kinematical twists, a likely signature of
triaxiality (e.g., NGC~1023).  The amplitude $c_{3}$ can be considered
as the \emph{morphology parameter}, as it encompasses most of the
`geometry' of the velocity field: third-order reconstructed fields
already display most of the interesting features.

\paragraph{Filtering.}
\label{sec:filter}
Since velocity maps are mostly point-antisymmetric ($V(r,\theta+\pi)
\simeq -V(r,\theta)$), the even moments will generally be smaller than
the odd ones. Going further, one can use kinemetry as a powerful
filtering tool to obtain the `closest' velocity field compatible with
a given intrinsic symmetry: a triaxial (bi-symmetric) galaxy has
$V(-x,-y) = -V(x,y)$, giving $a_{0} \equiv 0, c_{2n} \equiv 0$; and an
axisymmetric galaxy ($V(x,-y) = V(x,y)$) requires furthermore that
$\phi_{2n+1} \equiv \mathrm{PA}$ (see Fig.~\ref{fig:kinemetry}).

\begin{figure}
  \begin{center}
  \includegraphics[width=0.7\textwidth]{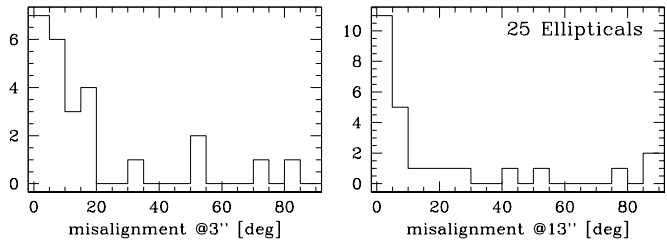}
  \end{center}
  \caption{Kinematic misalignment histogram for the 25~elliptical galaxies 
    of the \textsc{sauron} sample, measured at 3~arcsec (\emph{left panel})
    and 13~arcsec (\emph{right panel}).  }
  \label{fig:misalignment}
\end{figure}

\paragraph{Kinematic misalignment.}
\label{sec:misalignment}
The kinematic misalignment angle $\Psi$, defined as the difference between the
position angle of the major axis and the kinematic angle, is of particular
interest in the study of the intrinsic structure of early-type galaxies (e.g.,
Binney 1985; Franx \etal\ 1991).  Fig.~\ref{fig:misalignment} shows the
preliminary distribution of $\Psi$ in the representative sample of ellipticals
observed with \textsc{sauron} (de Zeeuw \etal, 2001). This can be compared
with Fig.~17 of Franx \etal{} (1991).

\section{Conclusions \& perspectives}
\label{sec:conclusion}

Kinemetry is a simple yet powerful and flexible technique to quantify
the kinematic maps of early-type galaxies as provided by IFS. It
allows filtering of the maps based on a chosen geometry of the galaxy.
The next step is to extend the analysis of, e.g., Statler
(1994), and work out from dynamical modeling how the various
kinemetric parameters -- in particular the misalignment angle $\Psi$
and the morphology parameter $c_{3}$ -- can be linked to the intrinsic
shape (triaxiality) and orientation of the observed galaxy, and thus
observationally constrain these quantities.


\end{document}